\RequirePackage{lineno}
\documentclass[aps,showkeys]{revtex4}

\usepackage{epsfig}
\usepackage{eepic}
\usepackage{latexsym}
\usepackage{rotating}
\usepackage{lscape}
\usepackage{graphics}
\usepackage{dcolumn}
\usepackage{bm}
\newcommand{\beq}{\begin{equation}}
\newcommand{\eeq}{\end{equation}}
\newcommand{\bd}{\begin{displaymath}}
\newcommand{\ed}{\end{displaymath}}

\newcommand{\bei}{\begin{itemize}}
\newcommand{\eei}{\end{itemize}}

\newcommand{\bee}{\begin{enumerate}}
\newcommand{\eee}{\end{enumerate}}

\usepackage[usenames]{color}
\usepackage{ulem}

\begin{document}

\noindent
\title{Impact  of cardio-synchronous brain pulsations on \\Monte Carlo calculated doses for synchrotron \\micro- and mini-beam radiation therapy} 

\author{Francisco Manchado de Sola}
\affiliation{%
Servicio de Radiof\'{\i}sica y Protecci\'on Radiol\'ogica, 
Hospital Juan Ram\'on Jim\'enez, Ronda Exterior Norte, s/n, E-21005 Huelva, Spain.}

\author{Manuel Vilches}
\affiliation{%
Servicio de Radiof\'{\i}sica y Protecci\'on Radiol\'ogica, Centro M\'edico de Asturias / IMOMA,
Avda. Richard Grand\'{\i}o, s/n, E-33193 Oviedo, Spain.}

\author{Yolanda Prezado}
\affiliation{Laboratoire Imagerie et Mod\'elisation en Neurobiologie et Canc\'erologie, 
CNRS, 15 rue Georges Clemenceau, F-91406 Orsay cedex, France.}

\author{Antonio M. Lallena}\thanks{Corresponding author}
\affiliation{Departamento de F\'{\i}sica At\'omica, Molecular y Nuclear, 
Universidad de Granada, E-18071 Granada, Spain.}

\date{\today}

\bigskip

\begin{abstract}
\noindent
{\it Purpose:} To assess the effects  of  brain movements induced by heartbeat on dose distributions in synchrotron micro- and mini-beam radiaton therapy and to develop a model to help guide decisions and planning for future clinical trials.\\
{\it Methods:} The Monte Carlo code PENELOPE was used to simulate the irradiation of a human head phantom with a variety of micro- and mini-beam arrays, with beams narrower than $100\,\mu$m and above $500\,\mu$m, respectively, and with radiation fields of $1\,$cm$\, \times\, 2\,$cm and $2\,$cm$\, \times\, 2\,$cm. The dose in the phantom due to these beams was calculated by superposing the dose profiles obtained for a single beam of $1\, \mu$m$\, \times\, 2\,$cm. A parameter $\delta$, accounting for the total displacement of the brain during the irradiation and due to the cardio-synchronous pulsation, was used to quantify the impact on peak-to-valley dose ratios and the full-width at half-maximum. \\
{\it Results:} The difference between the maximum (at the phantom entrance) and the minimum (at the phantom exit) values of the peak-to-valley dose ratio reduces when the parameter $\delta$ increases. The full-width at half-maximum remains almost constant with depth for any $\delta$ value. Sudden changes in the two quantities are observed at the interfaces between the various tissues (brain, skull and skin) present in the head phantom. The peak-to-valley dose ratio at the center of the head phantom reduces when $\delta$ increases, remaining above 70\% of the static value only for mini-beams and $\delta$ smaller than $\sim 200\,\mu$m. \\
{\it Conclusions:} Optimal setups for brain treatments with synchrotron radiation micro- and mini-beam combs depend on the brain displacement due to cardio-synchronous pulsation. Peak-to-valley dose ratios larger than 90\% of the maximum values obtained in the static case occur only for mini-beams and relatively large dose rates.
\end{abstract}


\maketitle

\section{\label{sec:intro} Introduction.}

Glioblastoma is the most frequent brain tumor in adult population, with 2-3 cases in one hundred thousand habitants per year\cite{NIH-stats}. It is a neuroectodermal malign neoplasm of the central nervous system, which shows a very fast growth, rapidly evolving to patient death, and that is considered by the World Health Organization as the most aggressive form of astrocity tumor \cite{WOH-GB}. Its treatment requires the delivery of high doses at an extended region around the tumor what may involve an unacceptable harm to healthy tissues \cite{Arraez2003}.

A possible strategy to overcome this limitation is to use a combination of radiation fields of submillimetric size with a spatial fractionation of the dose as in micro-beam (MRT) and mini-beam (MBRT) radiation therapies. MRT uses beams narrower than $100\, \mu$m with center-to-center (c-t-c) distances of 200 or $400\, \mu$m. The most commonly used combination and the one retained for future clinical trials is $50\, \mu$m-wide beams spaced by $400\, \mu$m \cite{Nettelbeck2009,Brauer2010,IMRovira2010,IMRovira2011}. MBRT involves beams with sizes above $500\, \mu$m and a c-t-c distance a factor two larger \cite{Dilmanian2006, Prezado2009}.  Contrary to conventional radiotherapy, the dose profiles in MBRT and MRT consist of peaks and valleys and high values of the peak-to-valley dose ratio (PVDR) are required for the specificity of the treatment: low valley doses in order to spare the normal tissues and high peak doses to achieve tumour control \cite{Dilmanian2002}. For given values of the peak width and the c-t-c, homogeneous irradiations of the tumor region can be obtained by using various incidence directions.

MRT and MBRT have been shown to significantly increase the normal tissue resistance: peak doses higher than $50\,$Gy, in one fraction irradiating the whole brain, are well tolerated by rat and mice brains \cite{Laissue2001, Dilmanian2006, Serduc2006,  Deman2012, Prezado2015}, in comparison with around 20 Gy in conventional seamless irradiation \cite{Calvo1988}. In addition, significant tumor growth delay in aggressive animal tumor models was observed \cite{Laissue2001, Dilmanian2006, Serduc2006,  Deman2012, Prezado2015}. Although a homogeneous tumor coverage would be preferred, numerous previous works in spatially fractionated techniques showed that tumor control can be obtained even with inhomogeneous dose distributions and high PVDR in the tumor \cite{Laissue1998, Serduc2009, Bouchet2010, Zwicker2004}. The tumoricidal effect of these techniques may be ascribed to the participation of some mechanism other than a direct death of tumor cells or the supression of their reproductive capacity, such as the induction of a denudation of the tumor vessel endothelium, tumor hypoxia \cite{Bouchet2013}, a decrease in tumor blood volume \cite{Bouchet2010, Bouchet2013}  or bystander effect/cellular communication   \cite{Crosbie2010}. These techniques could be also used as a boost, in addition to conventional radiotherapy treatment, aiming at increasing the dose in the tumor without dramatically enhancing the deleterious side effects. 

These promising radiotherapy approaches are nowadays being  explored at the European Synchrotron Radiation Facility (ESRF) \cite{Brauer2015}, the Brookhaven National Laboratory \cite{Dilmanian2006} and the Australian Synchrotron \cite{AusSync}.

However, one of the most important limitations to exploit MRT and MBRT in brain tumor treatments may be the blurring of the dose distributions due to cardio-synchronous pulsations. As the maximum speed reached by the brain due to heartbeat is $2\,$mm$\,$s$^{-1}$, extremely high dose rates are required to avoid the consequent beam smearing that would jeopardize the tissue sparing effect of the smallest micro-beams. In particular, at ESRF, dose rates above $5000\,$Gy$\,$s$^{-1}$ can be employed \cite{Brauer2005}. 

Despite this, a quantitative analysis of the effect of the cardio-synchronous pulsations and the brain movement they imply has not been carried out yet.
The aim of this work was to determine how that brain movement affects the dose distribution in MRT and MBRT. For this purpose we carried out Monte Carlo simulations of the brain irradiation for different configurations described in the literature as feasible therapy strategies. To do that we assumed a constant brain movement velocity and determined how this affects the photon beam profiles entering the irradiated organ. The analysis was done in terms of a parameter that provides the full organ displacement and is related to that velocity, the dose at the phantom entrance and the beam dose rate. By choosing the largest organ velocity found in the literature, a maximum limit for the effects of interest was evaluated. This may be used by MRT and MBRT practitioners to fix minimum dose rates and/or doses at the patient entrance in actual clinical situations or to determine the blurring level that can expected according to the specific dose and dose rate available at their facilities.

\section{\label{sec:matmet} Materials and methods.}

\subsection{Monte Carlo simulation}

Simulations performed in this work were carried out with the Monte Carlo code PENELOPE (v. 2014) \cite{salvat}. This is a general purpose code that permits to simulate the coupled transport of electrons, photons and positrons, with energies from 10~eV to 1~GeV, in any material geometry. Its ability to simulate in the low energy range makes this code particularly suitable for MRT and MBRT simulation.

Photon interactions are simulated in a detailed way. 
For electrons and positrons, PENELOPE makes use of a mixed simulation scheme, where interactions are classified as hard or soft. Hard events are simulated as photon interactions, one by one in a chronological way; all soft events between two hard interactions are grouped and described by means of a multiple scattering theory. The electron transport in each material is controlled by five parameters. $C_1$ controls the average angular deflection produced by the soft interactions between two consecutive hard events; $C_2$ is the maximum average fraction of energy lost between two consecutive hard interactions, and $W_{\rm CC}$ and $W_{\rm CR}$ define the threshold energies to discriminate between hard and soft interactions. Electron inelastic collisions in which the energy lost is bigger than $W_{\rm CC}$ and/or radiative interactions with an energy loss bigger than $W_{\rm CR}$ are considered hard events. A fifth parameter, $s_{\rm max}$, defines the maximum length that the electron can travel between two hard interactions.
In our simulations the values selected for these parameters as well as for the electron and photon absorption energies were $C_1 = C_2 = 0.05$, $W_{\rm CC}=5\,$keV, $W_{\rm CR}=10\,$keV, $E_{\rm abs}(e^-)=E_{\rm abs}(\gamma)=1\,$keV, and $s_{\rm max}$ was fixed to 1/10 of the characteristic dimension of each material element in the geometry, all of them within the range recommended in the user manual \cite{salvat}. PENELOPE has been used for dosimetry assessment in MRT \cite{IMRovira2010,IMRovira2011,IMRovira1,IMRovira2} and MBRT \cite{Prezado2009,prezado1,prezado2}.

\subsection{Beam configurations}

\begin{figure}[!b]
\begin{center}
\includegraphics[width=8cm]{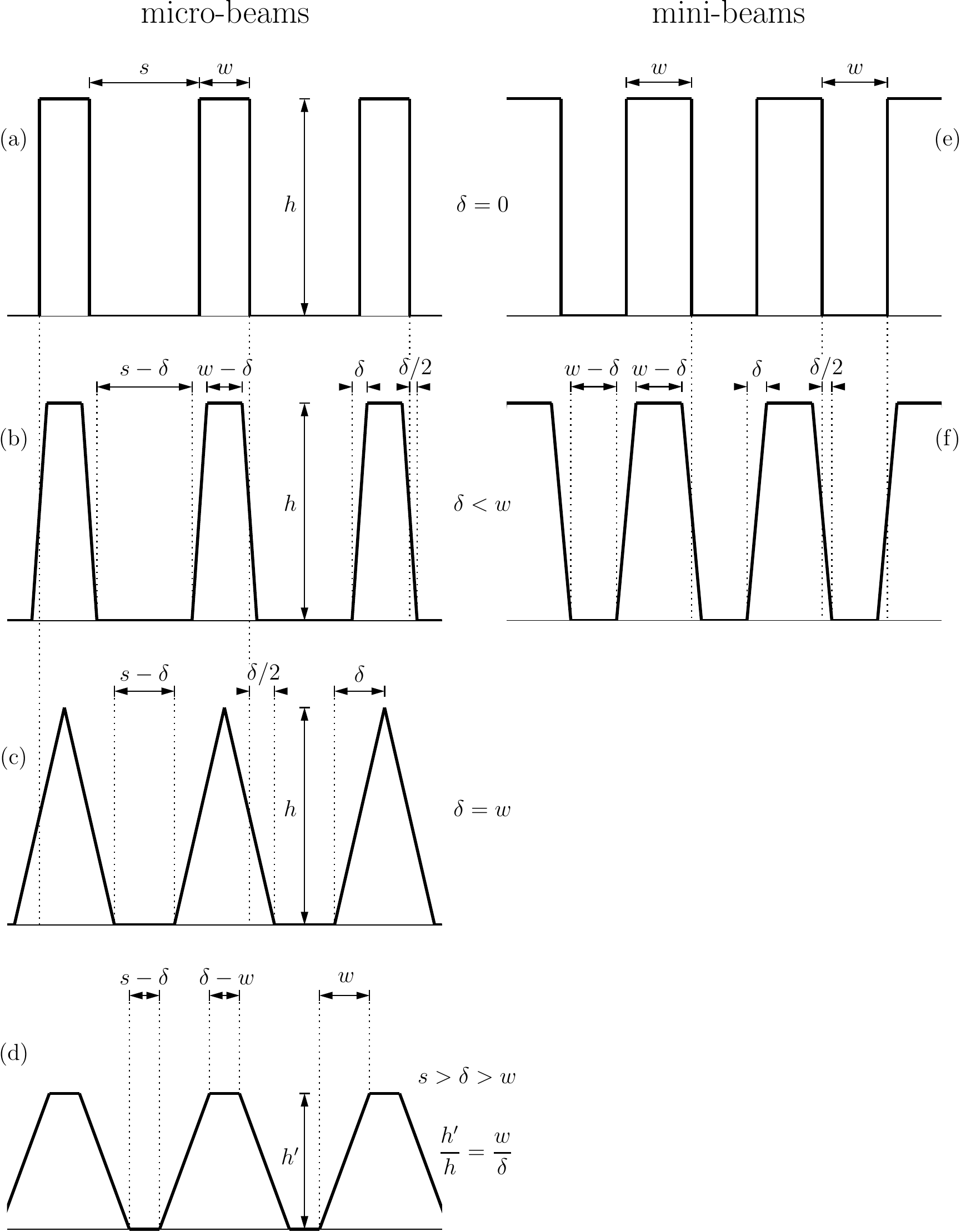}
\caption{\small Sketch of the photon fluences at the phantom entrance. Panels (a) and (e) show the actual incident fluence on the phantom for micro- and mini-beams, respectively. The other panels depict the apparent incident fluence seen by the moving brain in various situations characterized by the parameter $\delta$, which is defined in equation (\ref{eq:delta}) as the speed with which the target moves times the ratio of the total dose delivered and the dose rate. Note that micro-beams with $\delta > s$ are not considered because they would present overlapped peaks; for mini-beams only cases with $\delta \leq w$ have been analyzed.}
\label{fig:fluence}
\end{center}
\end{figure}

In this work photon parallel beams with radiation fields of size $1\,$cm$\, \times\, 2\,$cm and $2\,$cm$\, \times\, 2\,$cm were considered. They were micro- (figure \ref{fig:fluence}a) or mini-beams (figure \ref{fig:fluence}e) built up with several peaks that correspond to the regions of the dose profiles with the maximum dose. The peaks have a width $w$; the regions between two consecutive peaks are the ``valleys'', whose width is $s$. Thus, the c-t-c distance is given by $w+s$. Note that for mini-beams, $s=w$ and c-t-c equals $2w$.

It is worth mentioning that, in the simulations carried out, the irradiation fields were assumed to be produced as a whole, including the complete peak structure. In those synchrotrons where the height of the beam is limited to less than one mm \cite{prezado11,prezado1},  the irradiation is done by scanning the target with the help of a motorized platform driven by a high precision goniometer. The dose rates considered in our simulations are supposed to be measured at the target position, using the same scanning procedure, therefore taking into account, at least partially, that translation \cite{prezado11}. In any case, the convolution of the scanning motion of the target with that produced by heart beats would make the fluence seen by the target to have a rather complicated pattern that would require an analysis involving a very detailed description of the global target motion, including the irradiation details of the specific facility. With this in mind the results we quote here may be considered as a first, overall, indication of the effects that cardio-synchronous pulsations produce in this type of treatments. 

The energy spectrum of the photons entering the phantom is that of the ESRF ID17 Biomedical Beamline \cite{prezado1,IMRovira1}: it extends from about 50 to $600\,$keV, is peaked at $\sim 70\,$keV and has a mean energy of $\sim 100\,$keV.

To simulate the different configurations, we assumed that each mini- and micro-beam was formed by the superposition of the adequate number of adjacent and parallel 1~$\mu$m width beams. Then instead of simulating all the different beam configurations analyzed in the present work (see below), we simulated a single 1~$\mu$m beam with $10^9$ initial photons. The dose profiles corresponding to the actual mini- and micro-beams analyzed were calculated by numerical superposition of those obtained in that simulation. Typically, the beam configurations studied include about $10^4$ beams of 1~$\mu$m width and, as a consequence, the results obtained with our superposition procedure are equivalent to a simulation of the whole mini- and micro-beam with $10^{13}$ initial photons. This procedure permitted us an enormous saving of CPU time without requiring the use of variance reductions techniques to reach adequate statistics.

The verification of the superposition procedure was done by simulating the experimental setup described by Prezado et al. \cite{prezado2} and comparing the results obtained for PVDR and FWHM to those quoted by Prezado et al. \cite{prezado1,prezado2} Two reference mini-beams with 10 and 16 $w=600\,\mu$m peaks, respectively, were considered. The mini-beams impinged orthogonally onto the phantom surface. The phantom was a cylinder with $8\,$cm of radius and a height of $16\,$cm, filled with water. Scoring voxels of 1~$\mu$m~$\times$~2~cm~$\times$~1~mm were considered to determine the dose profiles that were calculated in the $x$ direction, with the beam traveling along the $z$ axis.

\subsection{Dosimetric parameters}

Throughout this work, we have used the peak-to-valley dose ratio, PVDR, and  the full-width at half-maximum, FWHM, to characterize the dose profiles at a certain depth in the phantom.

\begin{figure}[!th]
\begin{center}
\includegraphics[width=6cm]{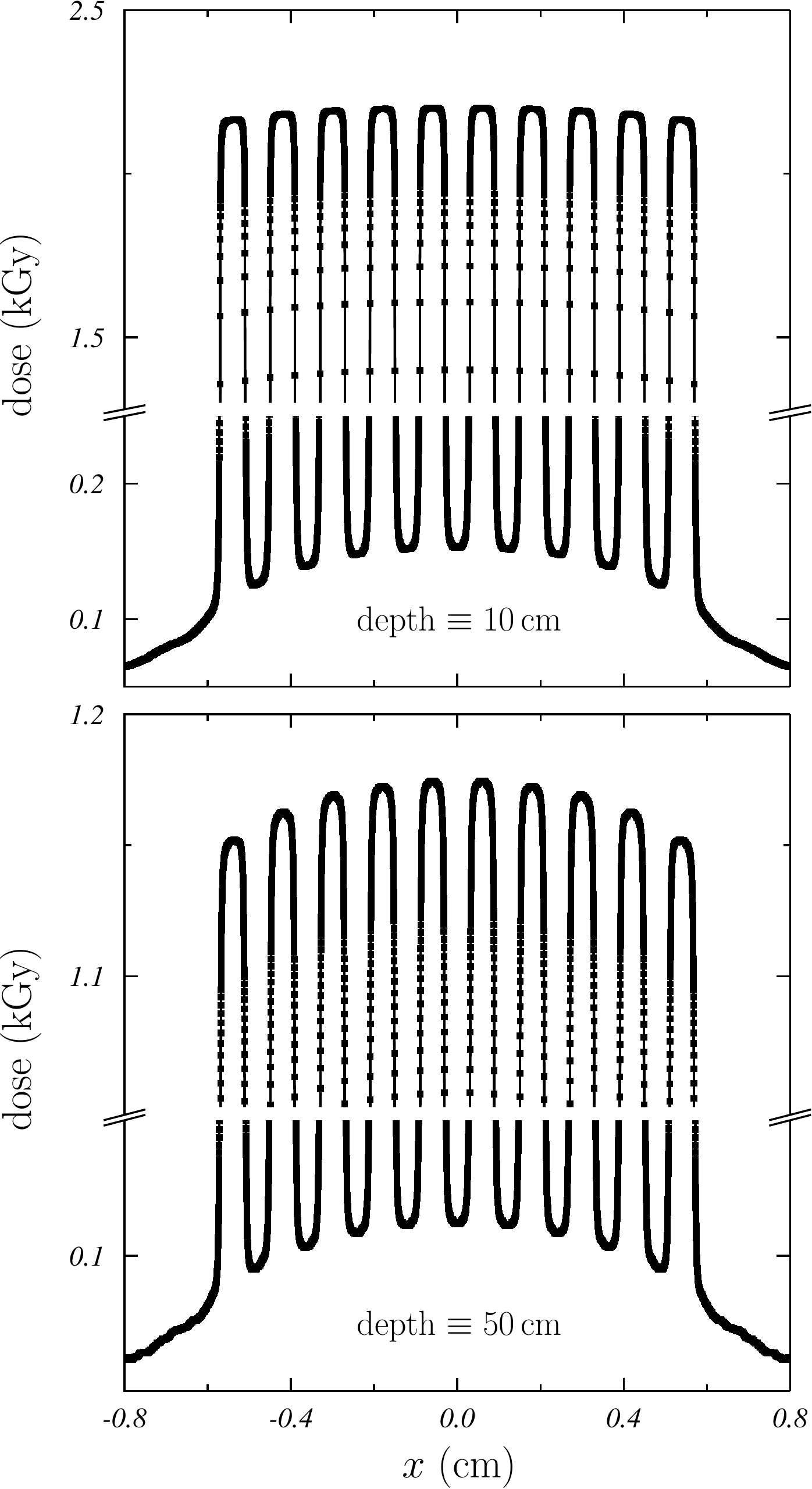}
\caption{\small Mini-beam dose profile at depths of $10$ and $50\,$cm in a water phantom.}
\label{fig:PROF}
\end{center}
\end{figure}

To calculate PVDR we proceeded as follows. First, for each peak in the profile, we determined the average peak dose $\bar{d}_{\rm p}(i)$ by averaging over a region of width $w$ centered around the peak center. As we had $n$ peaks in the radiation field we got $n$ average values $\{\bar{d}_{\rm p}(i),i=1,2,\ldots,n\}$. The same was done for the valleys, but considering a region of dimension $s$ centered at the center of each valley. As we had $n-1$ valleys in the radiation field this gave us $n-1$ average values $\{\bar{d}_{\rm v}(i),~i=1,2,\ldots,n-1\}$. Using these average values we calculate $n-1$ ``local'' PVDRs as ${\rm PVDR}_{\rm l}(i)=\bar{d}_{\rm p}(i)/\bar{d}_{\rm v}(i)$. Finally,
PVDR was calculated as the average of these local PVDRs:
\begin{equation}
{\rm PVDR} \,=\, \frac{1}{n-1} \,\sum_{i=1}^{n-1} {\rm PVDR}(i) \, .
\label{eq:PVDR}
\end{equation}
 As seen in figure \ref{fig:PROF}, both peaks and valleys values vary from the center to the sides of the beam. This produces a variation in the local PVDR values that is much larger than the statistical (type A) uncertainty due to the Monte Carlo calculations (which is of the order of $\sim 1\%$). For that reason we assumed that this variability represents a type B uncertainty of PVDR that was determined as the standard deviation of the local PVDRs. 

FWHM was determined as the average value of the FWHMs calculated for each one of the peaks in the profile. The corresponding uncertainties of FWHM were estimated in a way similar to that of PVDR.

From a biological point of view, PVDR is considered to be  a very relevant  dosimetric parameter \cite{Dilmanian2002,Brauer2005}. In fact, small values of PVDR imply a more uniform irradiation of healthy tissues and, therefore, a loss of the therapeutic advantage caused by the spatial fractioning characteristic of MRT/MBRT.

\subsection{Brain irradiation}

\begin{figure}[!bh]
\begin{center}
\includegraphics[width=3cm,angle=90]{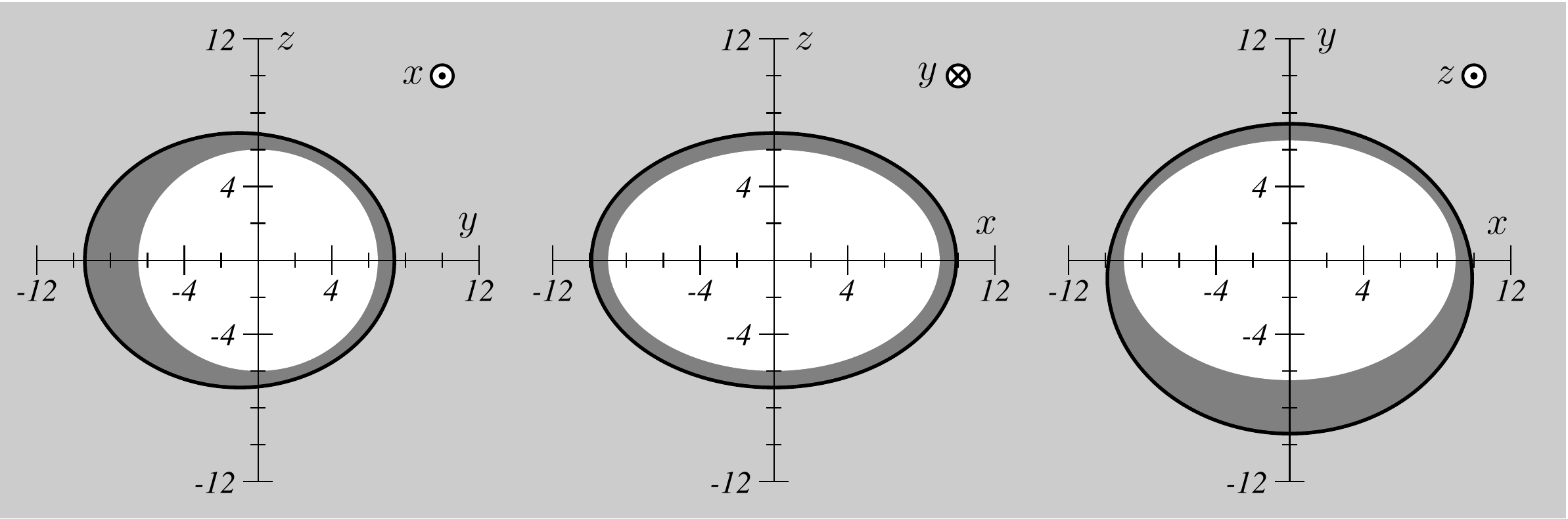}
\caption{\small Head phantom used in our simulations. Cuts at planes $x=0$ (left), $y=0$ (center) and $z=0$ (right) are shown. Dimensions are in cm. Skin (thin black area), skull (dark gray black) and brain (white) are shown. Air (clear gray) surrounds the phantom.}
\label{fig:headp}
\end{center}
\end{figure}

Dose distributions calculated in an anthropomorphic head phantom \cite{harling} were used to evaluate the effect of the cardio-synchronous brain pulsations. Figure \ref{fig:headp} shows the cuts at planes $x=0$ (left), $y=0$ (center) and $z=0$ (right) of this phantom. This is a mathematical model in which skin (thin black area), skull (dark gray) and brain (white) are described as ellipsoids. The phantom is surrounded by air and the source is situated at a distance of $1\,$m from the phantom surface also inside air. The materials used are those included in the PENELOPE database for air, skin, dense bone (for the skull) and brain.  

The reference system was centered at the geometrical center of the brain, with the $x$ axis oriented in the antero-posterior direction, the $y$ axis in cranio-caudal direction and the $z$ axis in the transverse direction. The scoring voxels used in the simulations of the $1\,\mu$m beam in this case were of 1~$\mu$m~$\times$~2~cm~$\times$~2~mm in size. The beam traveled in the $z$ positive direction and the profiles were determined in the $x$ direction. In this geometry both static and dynamic conditions were considered.

As said above, the mini-beam configurations considered had a c-t-c of $2w$, with $s=w$; on the other hand, in the micro-beams analyzed the valley width was $s=7w$, with a c-t-c of $8w$. 

In order to determine the effect of the brain movement, the dose rate, $\dot D$, the total dose delivered, $D$, and the speed with which the target moves, $v$, must be taken into account. In fact, the total displacement $\delta$ of the brain during the irradiation is given by
\begin{equation}
\delta \, = \, \frac{D}{\dot D} \, v \, .
\label{eq:delta}
\end{equation}
The moving target ``sees'' an apparent incident fluence as that sketched in figure \ref{fig:fluence}b. Each micro- and mini-beam is then modified: an initial ramp with a width $\delta$ is followed by a central, constant, region with a width equals to $w-\delta$ and a final descending ramp again with width $\delta$. The distance between the end of a peak and the beginning of the next one is $s-\delta$. Due to the movement there are brain regions that are actually irradiated only during a fraction of the total irradiation time. As $v$ is assumed to be constant, the effect of this movement can be described by modifying the beam in order to include the linear ramps shown.

In practice, to simulate the brain movement, the $x$ coordinate of the photon incident in the phantom was sampled according to this apparent fluence distribution. The total dose was reconstructed as in the static case, that is from the dose due to a beam of 1$\mu$m, but  taking into account now the variation of the fluence indicated by the trapezoidal distributions shown in figure \ref{fig:fluence}.

To investigate the effects of the brain movement on the various parameters of interest, we have considered dose rates between $5000$ and $30000\,$Gy/s for MRT, the latter being an extreme value that would be of interest in some applications and that could be reached in XFEL facilities \cite{XFEL}. In MBRT dose rates in the range $400-5000\,$Gy/s were analyzed. The doses at the phantom entrance may vary between $300$ and $600\,$Gy in the case of micro-beams \cite{Brauer2010}, while for mini-beams doses up to $100\,$Gy were considered \cite{Dilmanian2006, Prezado2009}. In what respect to the brain speed, a maximum $v=0.2\,$cm/s was quoted in reference \cite{Poncelet1992}. Thus, this value would be the most unfavorable and we assumed it in all calculations discussed below. The range of values considered for the parameter $\delta$ was between $40\,\mu$m and $500\,\mu$m, the latter being the maximum brain shift during the whole heart cycle. In the case of MRT, some of the largest values, $\delta=240$ and $500\,\mu$m, were not considered to avoid the overlap of consecutive beam peaks. The static case is recovered for $\delta=0$. The configurations analyzed for both MRT and MBRT are shown in table \ref{tab:config}.

\begin{table}[!t]
\caption{\small MRT and MBRT configurations analyzed. For each one the beam width $w$, the valley width, $s$, and the total displacement $\delta$, defined in equation (\ref{eq:delta}), are given. Also the number of peaks required to cover the whole radiation field is indicated.}
\begin{tabular}{rcccccc}
\hline\hline
 & field size & $w$ ($\mu$m) & $s$ ($\mu$m) & c-t-c ($\mu$m) &$\delta$ ($\mu$m) & \# of peaks \\ \hline
MRT & $1\,$cm$\, \times\, 2\,$cm & 25 & 175 & 200 & 0/40/80/120 & 50 \\
         &                                           & 50 & 350 & 400 & 0/40/80/120/240 & 25 \\ \cline{2-6}
         & $2\,$cm$\, \times\, 2\,$cm & 25 & 175 & 200 & 0/40/80/120 & 100 \\
         &                                           & 50 & 350 & 400 & 0/40/80/120/240 & 50 \\ \hline
MBRT & $1\,$cm$\, \times\, 2\,$cm & 600 & 600 & 1200 & 0/40/80/120/240/500 & 8 \\
         &                                           & 1000 & 1000 & 2000 & 0/40/80/120/240/500 & 5  \\ \cline{2-6}
         & $2\,$cm$\, \times\, 2\,$cm & 600 & 600 & 1200 & 0/40/80/120/240/500 & 16 \\
         &                                           & 1000 & 1000 & 2000 & 0/40/80/120/240/500 & 10 \\
\hline\hline
\label{tab:config}
\end{tabular}
\end{table}

\section{Results.}

\subsection{Reference simulations}

Table \ref{tab:tab1} shows the values we obtained for PVDR at different depths in water for beams including 10 and 16 peaks with $w=600\, \mu$m. Our results are statistically compatible with the experimental ones measured with radiochromic film by Prezado et al. \cite{prezado1,prezado2} The PVDR values obtained in the case of the beam with 10 peaks by these authors with simulations similar to ours are larger, $\sim 8\%$ at most, than those we have found with the superposition procedure. In any case, both results agree within the uncertainties. 

\begin{table}[!b]
\caption{\small PVDR values corresponding to two configurations of mini-beams including 10 and 16 peaks with a width of $600\, \mu$m each. The results obtained in our simulations are compared to those of previous works \cite{prezado1,prezado2} both simulated and measured with radiochromic film. Uncertainties are shown with a coverage factor $k=1$ (that is, a 68\% confidence interval). }
\footnotesize\rm
\begin{tabular}{ccccccccccc}
\hline\hline
 &   & \multicolumn{4}{c}{10 peaks} & &\multicolumn{4}{c}{16 peaks} \\ \cline{3-6} \cline{8-11} 
  &~~ &\multicolumn{2}{c}{PENELOPE} & ~~ &experimental & ~~ & \multicolumn{2}{c}{PENELOPE} & ~~ &experimental \\\cline{3-4} \cline{6-6} \cline{8-9} \cline{11-11}
  depth (cm)&&  superposition & Ref. \cite{prezado2} && Ref. \cite{prezado2} && superposition & Ref. \cite{prezado1} && Ref. \cite{prezado1}  \\ \hline 
  0.3 && $17.0\pm 0.5$ &&&&& $14.8 \pm 0.4$ & $14.04 \pm 0.07$ && $15.2\pm 1.6$ \\
  0.5 && $14.9\pm 0.4$ &&&&& $13.0 \pm 0.4$ & $12.24 \pm 0.06$ && $12.8 \pm 1.2$ \\
  1.0 && $12.4\pm 0.4$ &&&&& $10.7 \pm 0.3$ & $10.05\pm 0.05$ && $9.6\pm 1.0$ \\
  2.0&&  $10.4 \pm 0.3$ & $11.3 \pm 0.7$ && $10 \pm 2$ && $8.9\pm 0.3$ & $8.44\pm 0.04$ && $9.3\pm 1.0$ \\
 4.0 && $9.4 \pm 0.3$ & $9.6 \pm 0.6$ && $10 \pm 2$ && $7.8\pm 0.2$ & $7.38\pm 0.04$ && $8.6\pm 0.8$ \\
 6.0 && $8.9 \pm 0.3$ &&&&& $7.3\pm 0.3$ & $6.94\pm 0.04$ && $7.5\pm 0.8$ \\
 8.0 && $8.7 \pm 0.4$ & $8.9 \pm 0.5$ && $9.3 \pm 1.5$ && $7.1\pm 0.3$ & $6.67\pm 0.04$ && $6.7\pm 0.6$ \\ \hline
\hline
\label{tab:tab1}
\end{tabular}
\end{table}

The contrary occurs for the beam with 16 peaks where the PVDR values quoted in reference \cite{prezado1} are smaller (between 5 and 6\%) than ours. However, the uncertainties of the values found by Prezado et al. \cite{prezado1} are much smaller than those of our data. The reason for the large difference between the uncertainties of the results obtained in these two simulations may be ascribed to the fact that both peaks and valleys in the dose profile show a smooth gradient, the dose slightly reducing as one moves away from the center of the beam. As indicate above, this ``intrinsic variability'' has been included in our results and in those of reference \cite{prezado2}, but not in the calculation of reference \cite{prezado1} where the PVDR was estimated with the values in the central peak of the dose profile. 

\subsection{Brain irradiation}

\begin{figure}[!th]
\begin{center}
\includegraphics[width=9cm]{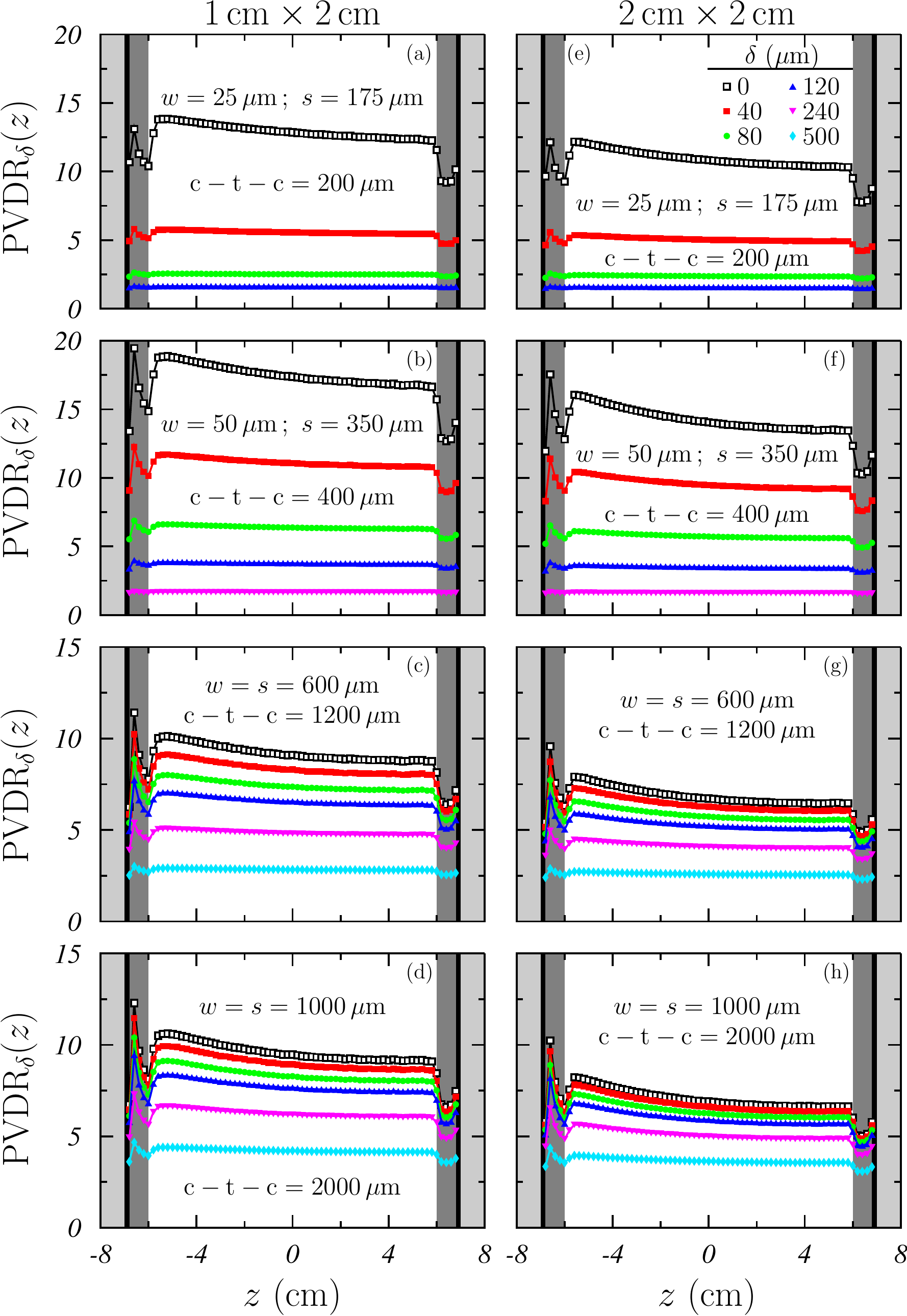}
\caption{\small PVDR$_\delta$ in the head phantom, as a function of $z$, for the various beam configurations analyzed (see table \ref{tab:config}). Air (clear gray), skin (thin black area), skull (dark gray) and brain (white) are indicated (as in Fig. \ref{fig:headp}). The different symbols correspond to the various $\delta$ values considered. Left (right) panels show the results obtained for a field of $1\,$cm$\, \times\, 2\,$cm ($2\,$cm$\, \times\, 2\,$cm).}
\label{fig:PVDR-1}
\end{center}
\end{figure}

Figure \ref{fig:PVDR-1} shows the variation with $z$ of the PVDR calculated for the different beam configurations analyzed (see table \ref{tab:config}). Left and right panels correspond to $1\,$cm$\, \times\, 2\,$cm and $2\,$cm$\, \times\, 2\,$cm fields, respectively. Apart from the sudden changes observed at the interfaces between the various tissues (brain, skull and skin) present in the head phantom, several aspects deserve to be pointed out. First, PVDR reduces as $\delta$ increases and this occurs at any depth in the phantom, independently of the head region (skin, skull or brain). 

\begin{figure}[!th]
\begin{center}
\includegraphics[width=6.7cm]{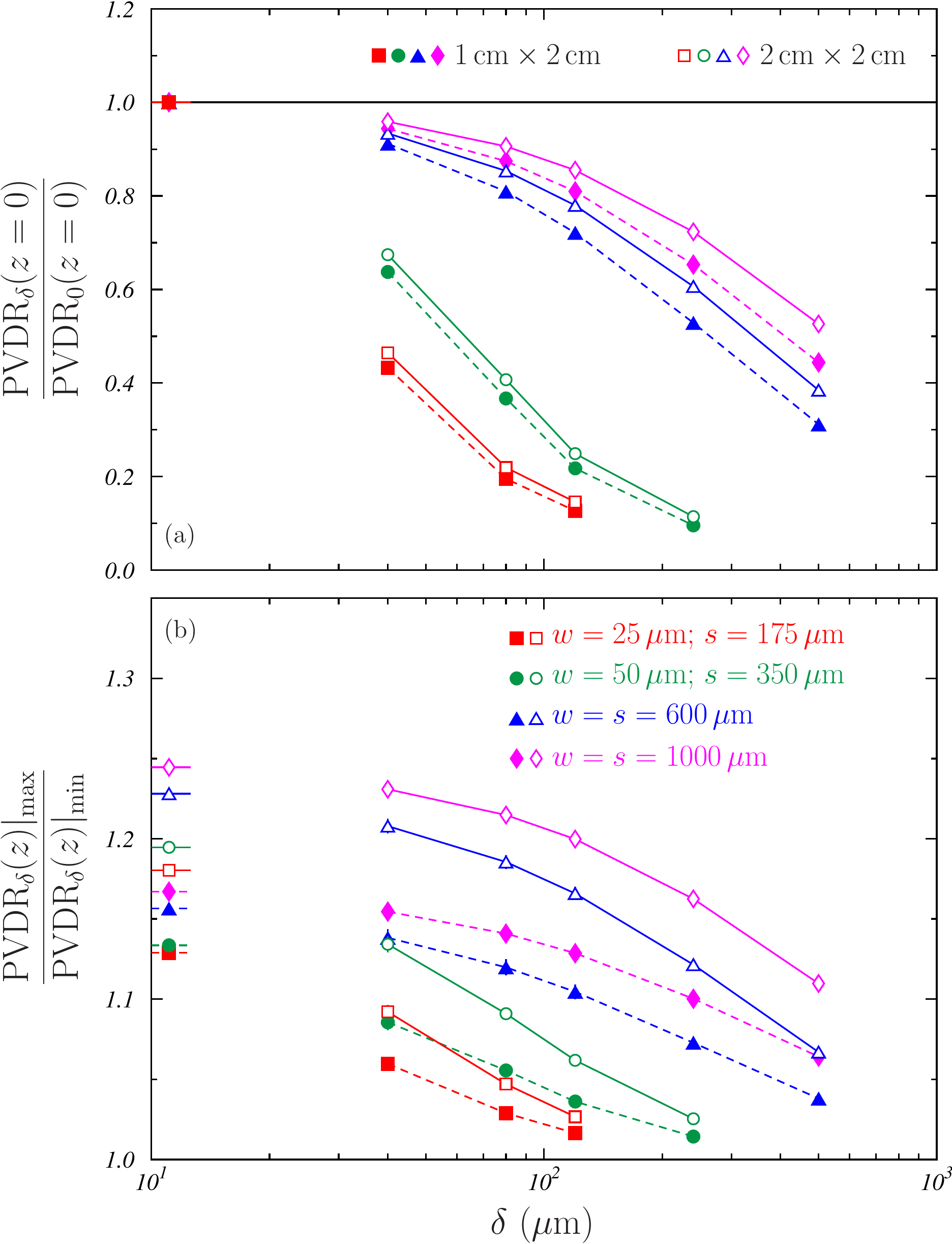}
\caption{\small Dependence with $\delta$ of (a) the ratio between PVDR$_\delta(z=0)$ and the corresponding value in the static case, PVDR$_0(z=0)$, and (b) the ratio between the maximum and the minimum values of PVDR$_\delta(z)$ found inside the brain region, for the various configurations analyzed (see table \ref{tab:config}). Solid and open symbols correspond to the results obtained for the $1\,$cm$\, \times\, 2\,$cm and $2\,$cm$\, \times\, 2\,$cm fields, respectively. The data points on the left edge of the panels are those found for $\delta=0$ and are shown as a reference; those in panel (a) are all of them 1.}
\label{fig:PVDR-z0}
\end{center}
\end{figure}

A more quantitative idea of this fact can be found from figure \ref{fig:PVDR-z0}a. Therein the ratio between PVDR$_\delta(z=0)$ and the corresponding value in the static case, PVDR$_0(z=0)$, is shown as a function of $\delta$ for the various configurations considered. It is worth noting that the ratios obtained for the $2\,$cm$\, \times\, 2\,$cm field are slightly larger than those corresponding to the $1\,$cm$\, \times\, 2\,$cm one. The reduction in PVDR within the brain is relatively smaller in the case of the largest field size or, in other words, the brain movement produces an effect smaller the larger the irradiation field is.

The effect of the brain movement is much more relevant in the case of the micro-beams. In fact, for $\delta=40$ we found $\displaystyle \frac{{\rm PVDR}_\delta(z=0)}{{\rm PVDR}_0(z=0)} \sim 0.7$ for the beam with $w=50\,\mu$m (circles) and $\sim 0.5$ for that with $w=25\,\mu$m (squares), while for the two mini-beam configurations the ratio is above 0.9. This is due to the fact that in all situations the increase of $\delta$ produces an enhancement of the dose in the valleys. However, the dynamic regime for mini-beams is that described in figure \ref{fig:fluence}f, while in the case of micro-beams the situation sketched in figure \ref{fig:fluence}d is reached for high $\delta$ values: in that case an additional reduction of the peak height occurs. It is important noticing that a percentage above 70\% of the PVDR value obtained in the static case is only found for mini-beams and $\delta$ values up to $\sim 200\,\mu$m; for micro-beams the PVDRs are below 70\% of the static value in all irradiation configurations analyzed. Here it is then important how $\delta$ compares to the dose spatial fractioning that can be estimated with $s$.

A second point deserving to be mentioned is that PVDR reduces with the depth inside the head phantom. This is so because the radiation scattered in the regions irradiated by the beam peaks reaches the initially non-irradiated valleys, increasing the dose. Figure \ref{fig:PVDR-z0}b shows, for the various configurations analyzed, how the ratio between the maximum and minimum values of PVDR$_\delta(z)$ found inside the head phantom varies with $\delta$. This ratio is larger for the $2\,$cm$\, \times\, 2\,$cm field than for the $1\,$cm$\, \times\, 2\,$cm one because the scatter radiation affects more the valleys for the largest field. On the other hand it diminishes when $\delta$ increases because the dose due to the scattered radiation, mainly that in the valleys, reduces more slowly than that in the peak regions. It is also seen how the differences between the ratios obtained for both radiation fields diminish when $\delta$ increases: as $\delta$ grows the difference in the amount of scattered radiation produced in each field becomes less significant compared to the effect of the brain movement.

The ratios obtained in the static case ($\delta=0$) are shown by the points in the left edge of figure \ref{fig:PVDR-z0}b. All the results found for the largest field (open symbols) are above those obtained for the smallest one (solid symbols) and increase with $w$. When $\delta \not=0$, the variation is smaller than in the static case because the increase of the scatter radiation in the valleys is dominated by the dynamic shift modeled with the fluence ramps.

\begin{figure}[!t]
\begin{center}
\includegraphics[width=9.cm]{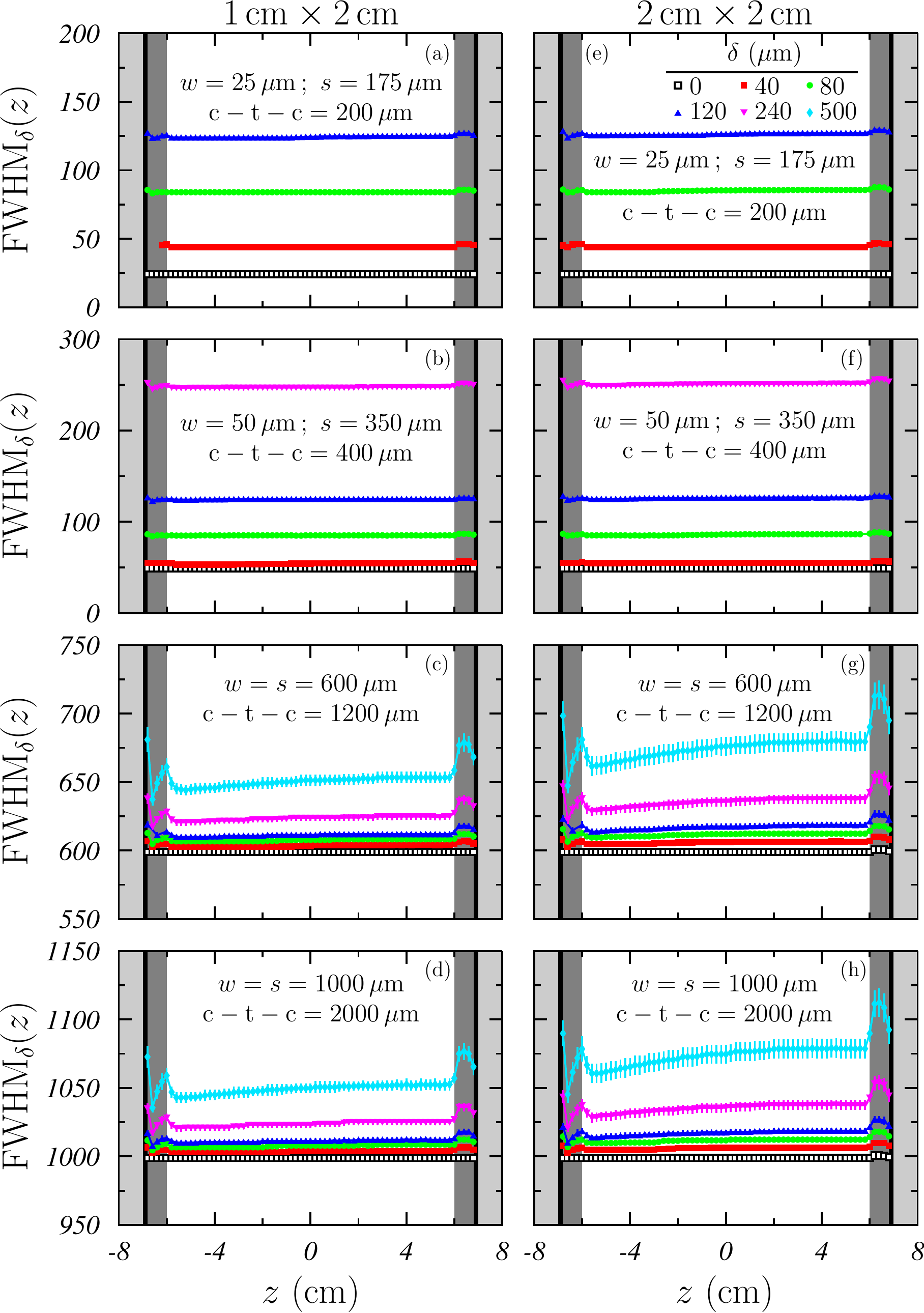}
\caption{\small FWHM$_\delta$ in the head phantom, as a function of $z$, for the various beam configurations analyzed (see table \ref{tab:config}). Air (clear gray), skin (thin black area), skull (dark gray) and brain (white) are indicated (as in Fig. \ref{fig:headp}). The different symbols correspond to the various $\delta$ values considered. Left (right) panels show the results obtained for a field of $1\,$cm$\, \times\, 2\,$cm ($2\,$cm$\, \times\, 2\,$cm).
}
\label{fig:FWHM-1}
\end{center}
\end{figure}

In figure \ref{fig:FWHM-1}, the variation of FWHM$_\delta$ with $z$ is shown for the configurations listed in table \ref{tab:config}. In all cases, FWHM$_\delta$ remains almost independent of the depth reached by the beam in the brain. In the case of the mini-beams considered (panels (c), (d), (g) and (h)), a slight enhancement as depth increases is observed for the configurations with the higher $\delta$ values. A similar trend was found also for the micro-beams, but it cannot be appreciated in panels (a), (b), (e) and (f) because of the scale of the figure. In summary, the variation of FWHM$_\delta$ with the depth in the brain is almost negligible in the static case and remains below 3\% for $\delta \not= 0$. This also makes the ratio $\displaystyle \frac{{\rm PVDR}_\delta(z)|_{\rm max}}{{\rm PVDR}_\delta(z)|_{\rm min}}$ to approach 1 as $\delta$ grows.

\begin{figure}[!th]
\begin{center}
\includegraphics[width=6.7cm]{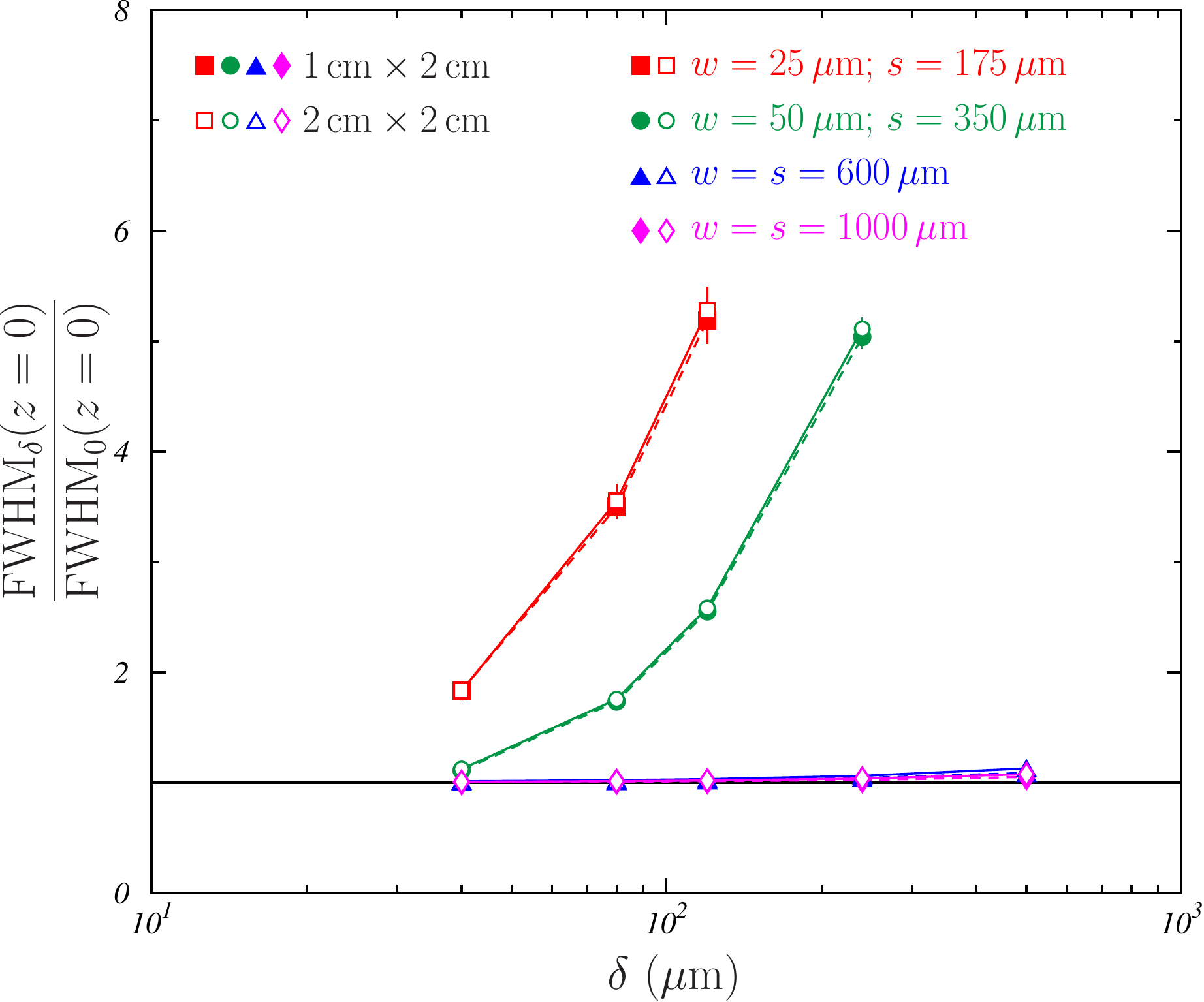}
\caption{\small Dependence with $\delta$ of the ratio between FWHM$_\delta(z=0)$ and the corresponding value in the static case, FWHM$_0(z=0)$, for the various configurations analyzed (see table \ref{tab:config}). Solid and open symbols correspond to the results obtained for the $1\,$cm$\, \times\, 2\,$cm and $2\,$cm$\, \times\, 2\,$cm fields, respectively.}
\label{fig:FWHM}
\end{center}
\end{figure}

Micro- and mini-beams show a different behavior of the ratio $\displaystyle \frac{{\rm FWHM}_\delta(z=0)}{{\rm FWHM}_0(z=0)}$. The results obtained for the beams studied are plotted in figure \ref{fig:FWHM} as a function of $\delta$. It can be seen that the ratio hardly varies with $\delta$, remaining practically equal to unity, in the case of the mini-beams (triangles and diamonds). However, in the case of micro-beams (squares and circles), FWHM$_\delta$ increases with $\delta$ and values up to 5 times those found in the static case are obtained. On the other hand, no differences are observed between the results corresponding to the two field sizes analyzed. This behavior is mainly due to the difference in the ratio $\delta/w$ between micro- and mini-beams. 

\begin{figure}[!h]
\begin{center}
\includegraphics[width=6.7cm]{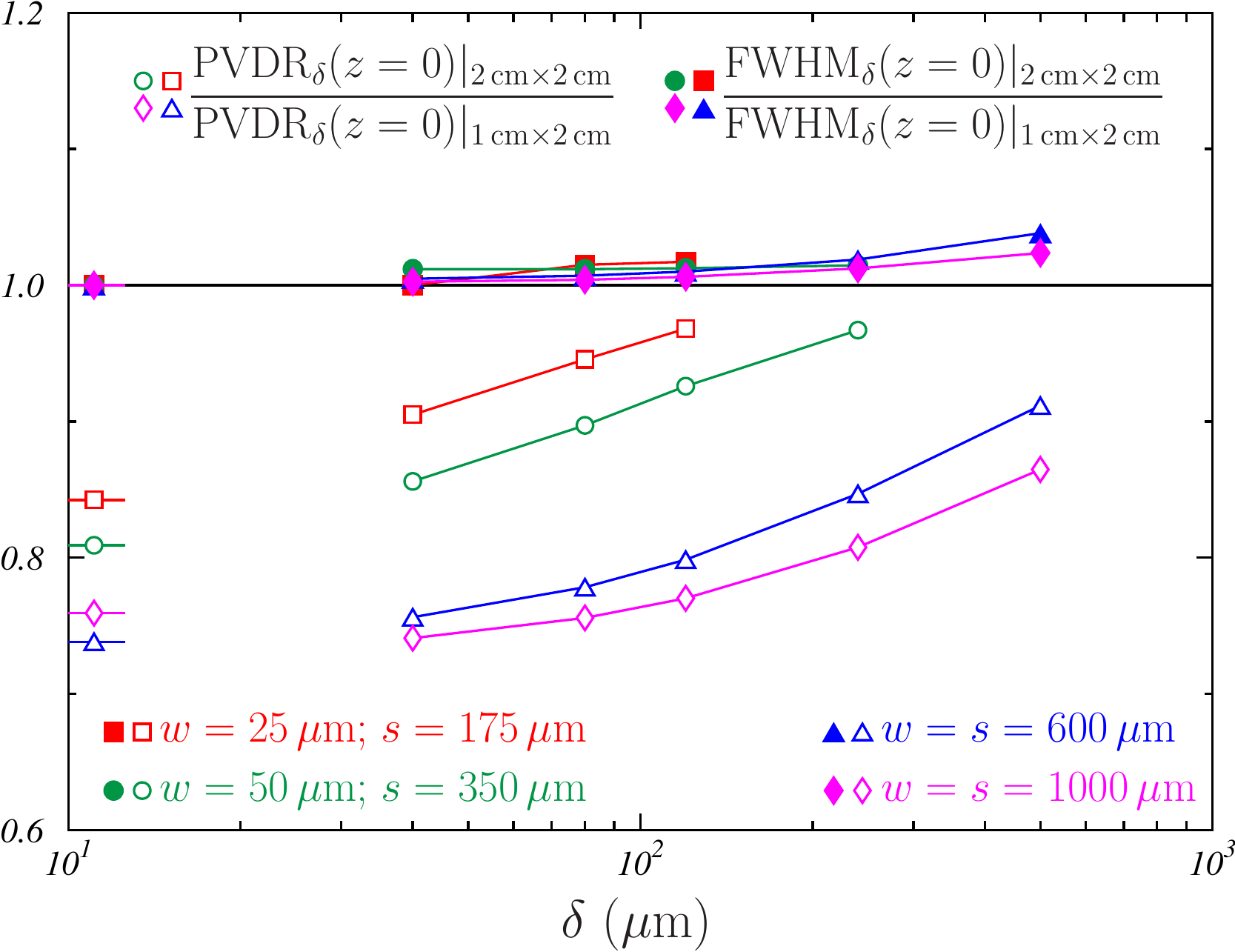}
\caption{\small Dependence with $\delta$ of the ratios $\displaystyle \frac{{\rm PVDR}_\delta(z=0)|_{2\,{\rm cm}\times 2\,{\rm cm}}}{{\rm PVDR}_\delta(z=0)|_{1\,{\rm cm}\times 2\,{\rm cm}}}$ (open symbols) and $\displaystyle \frac{{\rm FWHM}_\delta(z=0)|_{2\,{\rm cm}\times 2\,{\rm cm}}}{{\rm FWHM}_\delta(z=0)|_{1\,{\rm cm}\times 2\,{\rm cm}}}$ (solid symbols), for the various configurations analyzed (see table \ref{tab:config}). The data on the left edge are those found for $\delta=0$ and are shown as a reference.}
\label{fig:ratio}
\end{center}
\end{figure}

In figure \ref{fig:ratio}, the results found for the two field sizes considered are compared by means of the ratios $\displaystyle \frac{{\rm PVDR}_\delta(z=0)|_{2\,{\rm cm}\times 2\,{\rm cm}}}{{\rm PVDR}_\delta(z=0)|_{1\,{\rm cm}\times 2\,{\rm cm}}}$ and $\displaystyle \frac{{\rm FWHM}_\delta(z=0)|_{2\,{\rm cm}\times 2\,{\rm cm}}}{{\rm FWHM}_\delta(z=0)|_{1\,{\rm cm}\times 2\,{\rm cm}}}$. In the latter case (solid symbols), a very smooth enhancement is observed as $\delta$ increases, indicating that $2\,$cm$\, \times\, 2\,$cm beams widen slightly more than the $1\,$cm$\, \times\, 2\,$cm ones. The situation found for PVDR$_\delta$ is different (see open symbols). Though it also enhances as $\delta$ increases, the effect is much more pronounced than for FWHM$_\delta$. But the more important point is that the values are below 1, indicating that PVDR$_\delta$ is relatively larger in the case of the $1\,$cm$\, \times\, 2\,$cm field. In fact, for the static case (see data on the left edge of the figure) $\displaystyle \frac{{\rm PVDR}_\delta(z=0)|_{2\,{\rm cm}\times 2\,{\rm cm}}}{{\rm PVDR}_\delta(z=0)|_{1\,{\rm cm}\times 2\,{\rm cm}}}$ goes down to values $\sim 0.75$ for the two mini-beams considered (open diamond and triangle). The increase in the field size, going from $1\,$cm$\, \times\, 2\,$cm to $2\,$cm$\, \times\, 2\,$cm, enhances the scattered radiation: this enhancement modifies very slightly the peak widths (having almost no effect on FWHM), but increases significantly the dose in the valleys, thus reducing PVDR values. 

\begin{figure}[!th]
\begin{center}
\includegraphics[width=11cm,angle=0]{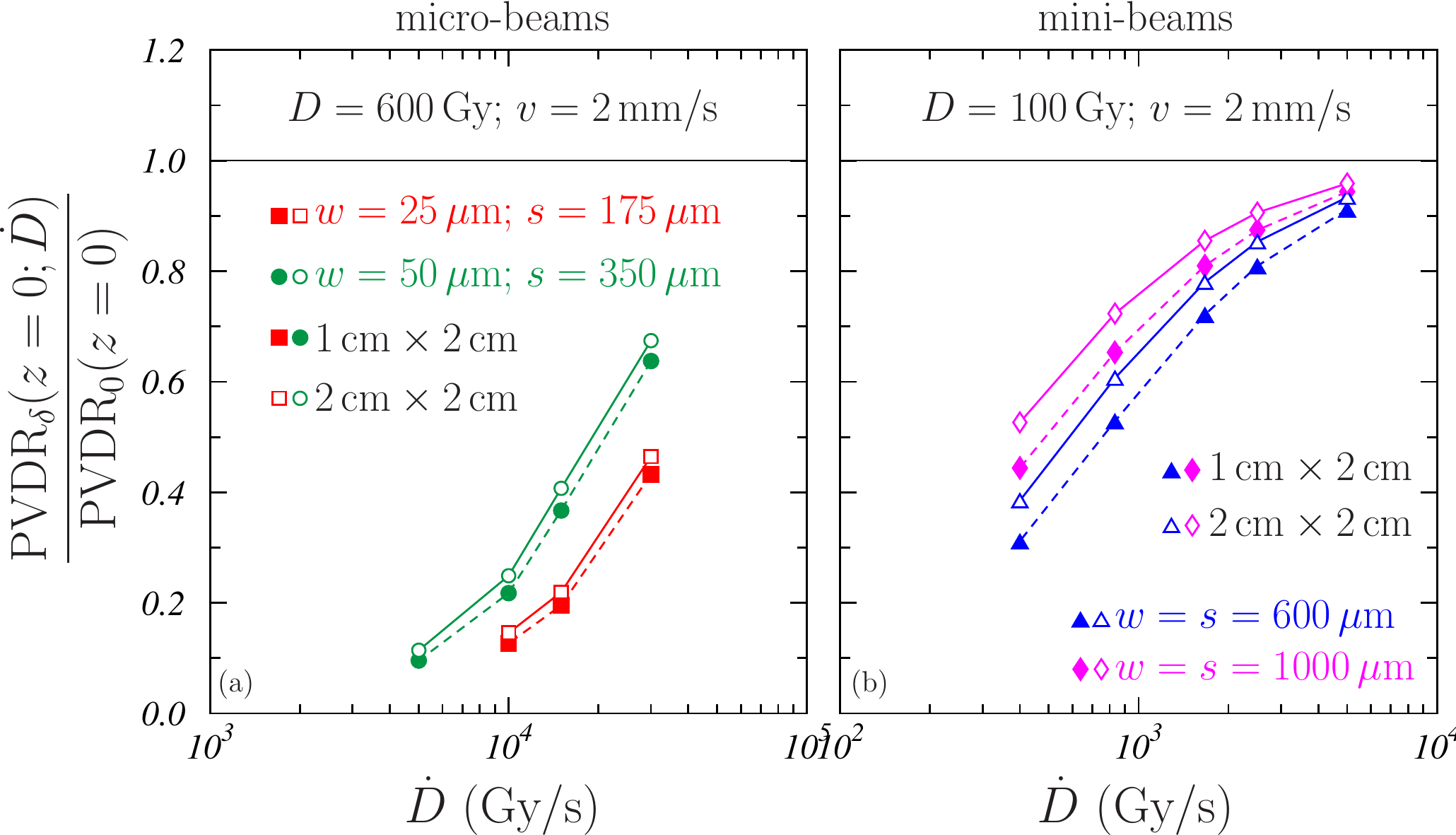}
\caption{\small Dependence with the dose rate of the ratio $\displaystyle \frac{{\rm PVDR}_\delta(z=0; \dot{D})}{{\rm PVDR}_0(z=0)}$, for various configurations. Solid and open symbols correspond to the results obtained for the $1\,$cm$\, \times\, 2\,$cm and $2\,$cm$\, \times\, 2\,$cm fields, respectively. Maximum doses at the entrance of the head phantom were assumed ($100\,$Gy for micro-beams and $600\,$Gy for mini-beams)}
\label{fig:rate}
\end{center}
\end{figure}

In figure \ref{fig:rate} the ratio $\displaystyle \frac{{\rm PVDR}_\delta(z=0; \dot{D})}{{\rm PVDR}_0(z=0)}$ is shown as a function of the dose rate for various configurations of micro- and mini-beams. As expected according to the results shown in figure \ref{fig:PVDR-z0}, and to the fact that $\delta$ and $\dot{D}$ are inversely proportional, the ratio grows with the dose rate in all cases. For mini-beams, shown in panel (b), values close to 1 are reached for the highest dose rates considered. For micro-beams (left panel) PVDR$_\delta$ is strongly affected: for dose rates of the order of $5000\,$Gy/s, only 50\% of the value obtained in the static case is recovered and smaller dose rates produce much smaller PVDR$_\delta$ values.

\begin{figure}[!th]
\begin{center}
\includegraphics[width=6.5cm,angle=90]{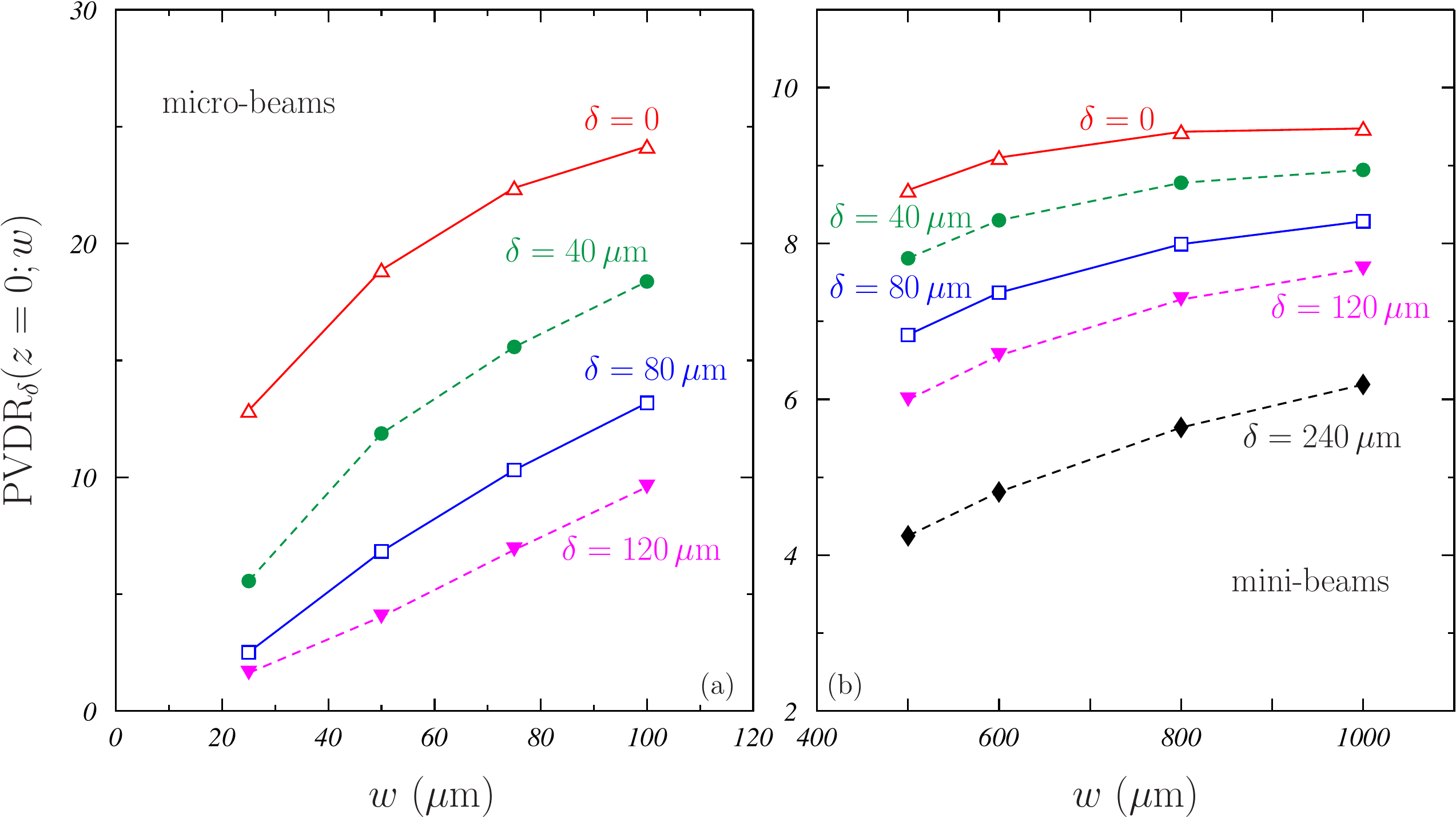}
\caption{\small Dependence with $w$ of ${\rm PVDR}_\delta(z=0;w)$ for a field of $2\,$cm$\, \times\, 2\,$cm. The static case ($\delta=0$) (red open triangles) is compared to those corresponding to various values of $\delta$. Results for micro- (panel (a)) and mini-beams (panel (b)) are shown.
}
\label{fig:www}
\end{center}
\end{figure}

Finally, we have analyzed how PVDR depends on the beam width $w$. In figure \ref{fig:www}, the values of ${\rm PVDR}_\delta(z=0;w)$ are shown versus this width for various $\delta$ values in the case of the $2\,$cm$\, \times\, 2\,$cm beams. To make the results comparable, the number of beam peaks included has been chosen in order to cover the whole field size. For micro-beams, values of $w$ ranging between $w=25\,\mu$m (50 peaks) and $w=100\,\mu$m (12 peaks) have been considered; for mini-beams $w$ was varied from $w=500\,\mu$m to $w=1000\,\mu$m, including between 10 and 5 peaks. In general PVDR$_\delta$ grows with $w$ but tends to saturate; this is more clearly seen in the case of mini-beams (right panel). In this respect it is worth emphasizing that, in the configurations investigated, both c-t-c and $s$ grow with $w$ (see table \ref{tab:config}).

On the other hand it is also apparent how PVDR$_\delta$ reduces as $\delta$ increases, reaching values about 60\% of those found for the static case in the limiting $\delta$ value, which corresponds to the maximum shift of the brain during the whole cardiac cycle.

\section{Conclusions.}
\label{sec:conc} 

The effect of the brain movement due to cardio-synchronous pulsation in the dosimetry of micro- and mini-beams of potential use in different radiotherapy treatments has been studied by means of Monte Carlo simulation with the code PENELOPE. The brain movement has been simulated by modifying the fluence of the beam incident onto the target. Specifically, the sharp limits of the peaks conforming the beams have been changed into linear ramps with a slope that depends on the total dose, the dose rate and the brain velocity through a parameter $\delta$ that represents the total displacement of the brain during irradiation ($\delta=0$ in the static case). Various beam configurations have been analyzed. 

An anthropomorphic head phantom has been considered to perform the dosimetry. The dose in the phantom has been reconstructed from the dose corresponding to a beam with a size of $1\, \mu$m$\, \times\, 2\,$cm and taking into account the corresponding beam fluences. The procedure has been checked by comparing the results obtained to those found for a complete beam and to other both experimental and simulated found in literature.

PVDR reduces as $\delta$ increases with the effect of the brain movement being more relevant for micro- than for mini-beams. It is also worth pointing out that PVDR shows a certain variability when it is calculated with the central peaks in the beam or with those situated at the extreme sides of it.

The variation of FWHM with the depth inside the brain region is very small, below 3\%.  for $\delta \not= 0$. However, while the values of FWHM remain practically identical to those of the static case for mini-beams, grow with $\delta$ for micro-beams, reaching values five times larger than those found for $\delta=0$. This is due to the relatively large value of $\delta$ with respect to $w$ in micro-beams.

The PVDR values obtained for the $1\,$cm$\, \times\, 2\,$cm field are larger than those found for the $2\,$cm$\, \times\, 2\,$cm one: for low $\delta$ values the corresponding ratios are around 0.75 for mini-beams and above 0.8 for micro-beams. However, the FWHM are essentially the same for both field sizes. PVDR grows when the peak width $w$ increases and tends to saturate. However, for a given $w$ value, it reduces as $\delta$ increases.

PVDR increases with the dose rate for a given total dose value. In case of mini-beams, assuming the maximum total dose analyzed $D=100\,$Gy, the values found for the static case are recovered at a 80\% for dose rates of the order of $1000\,$Gy/s. In the micro-beams studied, 60\% of the static PVDR values at most is reached for the highest dose rates analyzed and $D=600\,$Gy. 

In the calculations done in the present work, the brain velocity was chosen to be transverse to the direction of the peak and valley pattern of the beam and with its maximum published value \cite{Poncelet1992}. In this way we estimated the largest effects that cardio-synchronous brain movements may produce in the doses due to micro- and mini-beam radiotherapy. In any case, specific calculations for the particular cases of interest would be required to obtain quantitative detailed estimations of such doses. 

Doses may be affected significantly by the cardio-synchronous pulsation in case of MRT and the application to humans of his technique would require the use of cardiac gating. On the other hand, the tumor control could be compromised if very high peak doses, well above $300\,$Gy, are not used \cite{Serduc2009}. This implies the need to carry out biological studies that allow establishing the quantitative extent of the effects analyzed here.

MBRT doses are much less affected by the brain movement than MRT ones. In addition, lower doses seem to be required for a significant  tumor control \cite{prezado2,Deman2012}. Then feasible brain treatments could be carried out with dose rates smaller than in MRT thus reducing patient safety requirements. This also makes more viable the transfer of the technique towards conventional equipments. In the case of very high dose rates, rather stricter requisites such as, {\it e. g.}, those permitting to interrupt the irradiation in nanoseconds, may imply significant technical or engineering challenges.

The results our study may help MRT and MBRT clinicians to determine the optimal treatment parameters according to the capabilities of their facilities. Besides, our findings may be extrapolated to other anatomical regions with different movement velocities in a straightforward way.

\acknowledgments{This work has been supported in part by the
Junta de Andaluc\'{\i}a (FQM0387), the European Regional  Fund (ERDF) and by the Spanish Ministerio de Ciencia y Competitividad (FPA2015-67694-P).}

\vspace{1cm}

{\bf \small Conflict of interest}

The authors have no conflicts to disclose.

\newpage

\end{document}